%% file: main.tex
\documentclass[runningheads]{llncs}
\usepackage{algorithm}
\usepackage{algorithmic}
\usepackage{amsmath}
\usepackage{booktabs}
\usepackage{csquotes}
\usepackage{graphicx} 
\usepackage{hyperref}
\usepackage{subcaption} 
\usepackage{xcolor}
\usepackage{multirow}
\newcommand{\repthanks}[1]{\textsuperscript{\ref{#1}}}
\makeatletter
\patchcmd{\maketitle}
  {\def\thanks}
  {\let\repthanks\repthanksunskip\def\thanks}
  {}{}
\patchcmd{\@maketitle}
  {\def\thanks}
  {\let\repthanks\@gobble\def\thanks}
  {}{}
\newcommand\repthanksunskip[1]{\unskip{}}
\makeatother

\begin{document}
\title{How to Surprisingly Consider Recommendations? A Knowledge-Graph-based Approach Relying on Complex Network Metrics}
\label{concept-kg-based-recos}
\titlerunning{How to Surprisingly Consider Recommendations? A KG-based Approach.}
%
%
\author{Oliver Baumann\inst{1}\thanks{All authors contributed equally\protect\label{X}}\orcidID{0000-0003-4919-9033} \and Durgesh Nandini\inst{1}\repthanks{X}\orcidID{0000-0002-9416-8554} \and Anderson Rossanez\inst{2}\repthanks{X}\orcidID{0000-0001-7103-4281} \and Mirco Schoenfeld\inst{1}\orcidID{0000-0002-2843-3137} \and Julio Cesar dos Reis\inst{2}\orcidID{0000-0002-9545-2098}}

\authorrunning{Baumann et al.}
%
\institute{University of Bayreuth, Bayreuth, Germany\\ \email{\{oliver.baumann,durgesh.nandini,mirco.schoenfeld\}@uni-bayreuth.de} \and Institute of Computing, University of Campinas, Campinas, SP, Brazil\\ \email{\{anderson.rossanez,jreis\}@ic.unicamp.br}}
\maketitle              
%
\begin{abstract} 
\input{sections/abstract}

\end{abstract}
%
%
%
\input{sections/01_introduction}
\input{sections/02_related_work}

\input{sections/03_method}
\input{sections/04_evaluation}

\input{sections/05_discussion}
\input{sections/06_conclusion}

\section*{Supplemental Material Statement}
Source code and data for the experiments and evaluations conducted in this work are available at \url{https://anonymous.4open.science/r/kg-recommender-273D/}.
The LFM-1b dataset is available at \url{http://www.cp.jku.at/datasets/LFM-1b/}, the CultMRS dataset at \url{https://zenodo.org/records/3477842}, and the Netflix titles dataset at \url{https://www.kaggle.com/datasets/shivamb/netflix-shows}.

\begin{credits}
    \subsubsection{\ackname}
    This article is the outcome of research conducted within the Africa Multiple Cluster of Excellence at the University of Bayreuth, funded by the Deutsche Forschungsgemeinschaft (DFG, German Research Foundation) under Germany’s Excellence Strategy – EXC 2052/1 – 390713894. 
\end{credits}

%
\bibliographystyle{references/splncs04}
\bibliography{references/references, references/kg_recos}
%
\end{document}

%% file: sections/abstract.tex
Traditional recommendation proposals, including content-based and collaborative filtering, usually focus on similarity between items or users. Existing approaches lack ways of introducing unexpectedness into recommendations, prioritizing globally popular items over exposing users to unforeseen items. This investigation aims to design and evaluate a novel layer on top of recommender systems suited to incorporate relational information and suggest items with a user-defined degree of surprise. We propose a Knowledge Graph (KG) based recommender system by encoding user interactions on item catalogs. Our study explores whether network-level metrics on KGs can influence the degree of surprise in recommendations. We hypothesize that surprisingness correlates with certain network metrics, treating user profiles as subgraphs within a larger catalog KG. The achieved solution reranks recommendations based on their impact on structural graph metrics. Our research contributes to optimizing recommendations to reflect the metrics. We experimentally evaluate our approach on two datasets of LastFM listening histories and synthetic Netflix viewing profiles. We find that reranking items based on complex network metrics leads to a more unexpected and surprising composition of recommendation lists.

%% file: sections/01_introduction.tex
\section{Introduction}
\label{sec:introduction}

Recommender Systems (RSs) aim to offer a personalized view of large complex spaces, prioritizing items likely to interest the user by analyzing user preferences, historical behavior, and item characteristics \cite{felfernig2008constraint}. Recommendations can expose users to relevant items and expand their understanding of the catalog, regardless of whether in an e-commerce, media-streaming, or GLAM setting. The most popular approaches for recommender systems are collaborative filtering and content-based filtering \cite{schafer2007collaborative,sarwar2001item}. User-item recommendations are an important part of the discovery process of large collections. In content-based filtering, item characteristics are used to determine the similarity between items rated (viewed, listened, bought, etc.) by a user, and ``unseen'' items. Collaborative filtering, on the other hand, determines users similar to the target user and predicts ratings on unseen items by the target user.

While existing approaches have been shown to produce meaningful recommendations, the items they recommend tend to be expected and located in whatever portion of the catalog is considered ``mainstream''. These approaches do not consider the rich relations between items beyond the realm of similarity alone. We argue that users may profit from recommendations that present a surprise element, as they may come in touch with concepts they have been unaware of.

Knowledge Graph RSs combine the capabilities of RSs and Knowledge Graphs (KGs) by incorporating and analyzing the structured representation of information in KGs. These systems leverage the interconnected nature of entities and their attributes within the KG to enhance the accuracy and relevance of recommendations.
Using KGs, RSs can go beyond simple user-item interactions and incorporate a broader understanding of the relationships among items, users, and other entities. This allows for more sophisticated recommendation approaches that consider not only the user's preferences. In this sense, contextual information encoded in KGs influences recommendation items. For example, in a movie recommendation scenario, a KG-based recommender system could consider not only the user's past viewing history and ratings. It can consider, for instance, the genre of the movie, the actors and directors involved, and the relationships between movies based on shared themes or motifs.

In this study, we propose a layer on top of recommender systems, extending their functionality by a configurable degree of surprise. Our approach considers relational information among items encoded in KGs and suggests items with a user-defined degree of surprise relying on results generated by a recommender system. The main research question guiding our investigation is whether network metrics computed on the KG influence the degree of surprise within the recommendations. We propose taking advantage of KG's graph structure, employing complex network measurements \cite{Rossanez2023} as a resource for encoding entity relevance in a KG. Centrality measurements denote different meanings of relevance for graph nodes, bringing novelty aspects for analyses over KGs. Our assumption highlights that the ``surprisingness'' of recommendations is reflected in the network-level metrics of the KG.

Figure~\ref{fig:system-overview} provides a high-level overview of our approach. We construct KGs from two distinct catalogs: users' listening events on the platform LastFM\footnote{\url{https://www.last.fm/}}, and TV shows and movies on Netflix\footnote{\url{https://www.netflix.com}}.   User profiles for LastFM are available through the LFM-1b dataset\cite{Schedl2016}; for Netflix, we generate synthetic profiles. Recommendations for these profiles are then generated through state-of-the-art recommender systems (RS). Our work supports any RS, as we focus on re-ranking recommendations to surface surprising results.
Consequently, a specific RS optimal for a particular use case can be selected. For each user profile, we determine the induced subgraph on the catalog-KG that includes all items the user interacted with and further entities that enrich the model. Then, for each user and each item in their recommendation list, we assess the impact of including that item and its KG-informed neighborhood on the user's subgraph through pre-determined graph metrics. The original recommendation lists are then re-ranked according to their relative impact.

Our contributions are summarized as follows:
\begin{itemize}

\item Insert a configurable level of surprise to any recommender system by adding a layer of meta-analysis on obtained recommendations;

\item Identify a network metric that correlates with different dimensions of surprise;

\item Provide a comparative study regarding several network-level metrics for reranking recommendation results;

\end{itemize}

The remainder of this article is organized as follows: Section \ref{sec:related_work} discusses related work. Section \ref{sec:proposal} presents our proposal. Section \ref{sec:evaluation} reports our experimental evaluation and its results. Section \ref{sec:discussion} discusses our findings. Section \ref{sec:conclusion} wraps up our investigation and points out directions for future studies.

%% file: sections/02_related_work.tex
\section{Related Work}
\label{sec:related_work}

Joseph \& Jiang \cite{joseph2019content} proposed a graph traversal algorithm novel weighting scheme for cold-start content-based recommendation by utilizing named entities. Their work computes the shortest distance between named entities over large KGs. Wang \textit{et al.} \cite{wang2019kgat} introduced the KG Attention Network, which enhances the effectiveness of collaborative filtering in RSs by effectively modeling the high-order connectivity between users, items, and entities within a KG. Their research investigated how different levels of connectivity, first-order, second-order, third-order, etc. impact the model's effectiveness. They discussed the findings of using attention mechanisms and KG embeddings. 

Hui \textit{et al.} \cite{hui2022personalized} presented a RS called ReBKC that uses auxiliary information such as historical user behavior and KGs to provide personalized suggestions. Their investigation integrates KG embeddings and user-item interaction data to address issues like sparse data and cold starts. ReBKC suggests using KGs as heterogeneous networks to incorporate additional information to unify embeddings of user behavior and knowledge features.
Their proposed algorithm employs collaborative filtering, enhanced by the rich semantic associations in KGs, to mine user preferences more deeply. The system learns from historical user interactions and multiple types of relationships within the KG.

Zhang \textit{et al.} \cite{Zhang2016} addressed the limitations of collaborative filtering in RSs by leveraging heterogeneous information in a knowledge base to improve the quality of RSs. Their proposed framework -- Collaborative Knowledge Base Embedding (CKE) -- comprises three components to extract semantic representations from the structural, textual, and visual content of items. These components employ techniques such as heterogeneous network embedding, stacked denoising auto-encoders, and stacked convolutional auto-encoders to extract textual representations and visual representations. It then jointly learns the latent representations in collaborative filtering as well as items' semantic representations from the knowledge base. Kaminskas and Bridge~\cite{kaminskas2016diversity} looked into the aspects of diversity, serendipity, novelty, and coverage and explained that introducing surprise in RS can burst the \enquote{user filter bubble} by finding interesting items that the user might not have otherwise discovered.

Kotkov \textit{et al.} \cite{kotkov2016survey} examined the concept of serendipity in the context of RS. Their work discussed different approaches to measure and enhance serendipity in RS, including the use of algorithms that utilize uncommon similarity measures or adapt based on user feedback. Their investigation looked at the balance between accuracy and novelty in recommendations and explored both offline and online evaluation strategies for assessing the effectiveness of RS in delivering serendipitous results. On the other hand, De Gemminis \textit{et al.} \cite{de2015investigation} proposed to produce serendipitous suggestions by utilizing the knowledge infusion process. Their investigation addressed the overspecialization issue in RSs, proposing to enhance serendipity by suggesting surprising items. Their approach enrichs a graph-based recommendation algorithm with background knowledge to uncover hidden correlations among items. 

Baumann and Schoenfeld~\cite{Baumann2022} used a KG-based RS to evaluate diversity and novelty of recommendations, both on a content- and a network-level. Using subgraphs constructed from user profiles, they generated recommendations by favoring unpopular items in the catalog that exhibit a high distance from a user's profile in terms of content-based features. Apart from unexpectedness and diversity on a content level, they found this approach to result in a more fair degree distribution on the individual profile subgraphs.

To the best of our knowledge, our present study is the first to apply complex network measurements to rerank the order of RS results. Our approach looks at the graph structure within the KG changes to compute the metrics for obtaining surprising recommendations.

%% file: sections/03_method.tex
\section{KG-Informed Recommendation (Re-)Ranking}
\label{sec:proposal}

\begin{figure}[!htp]
\includegraphics[width=\textwidth]{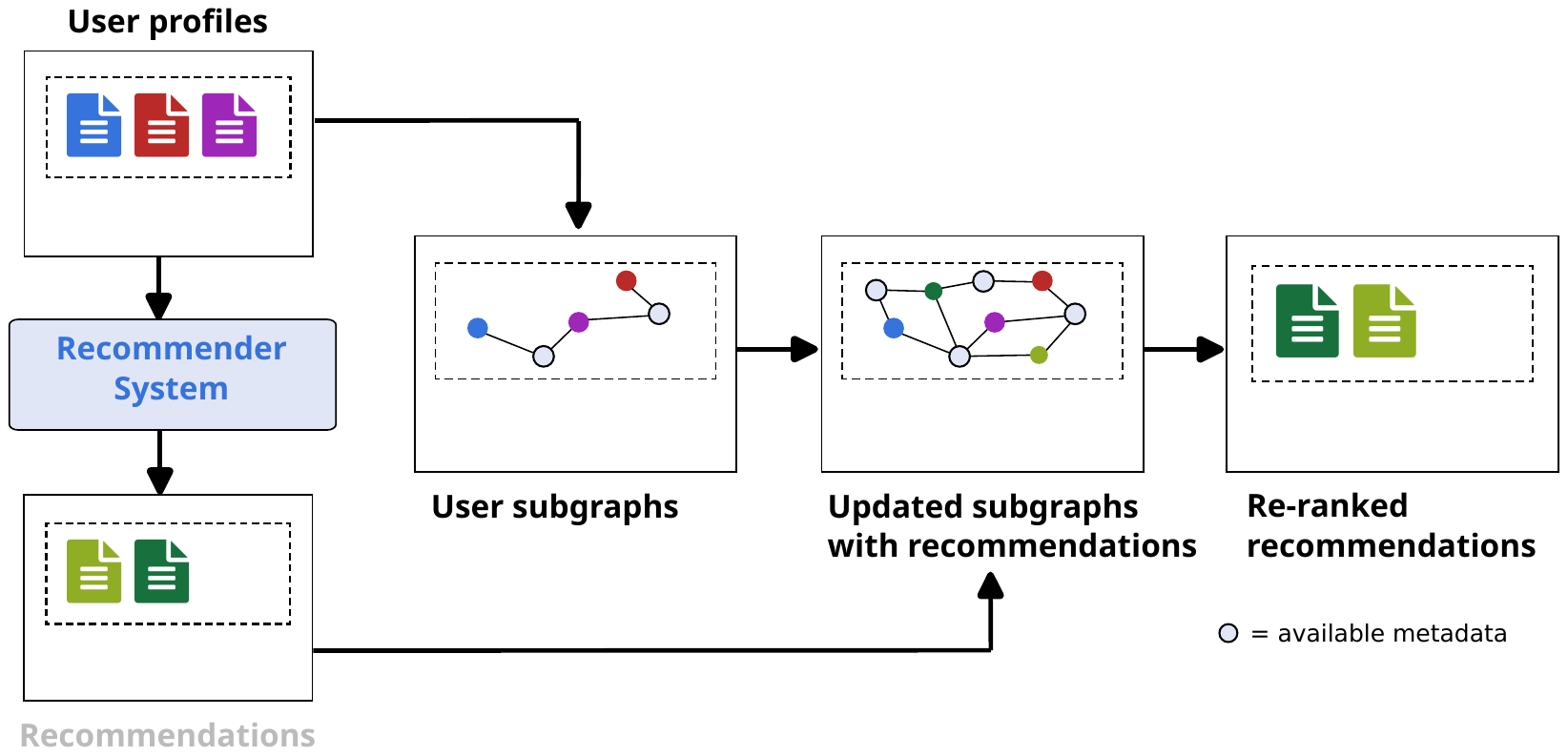}
\caption{Overview of the proposed knowledge-graph informed recommendations. KGs are constructed for the item catalog and all user profiles. The latter serve as input to an arbitrary state-of-the-art RS, whose results are re-ranked according to the impact the items would have were they included in the original user profile.} \label{fig:system-overview}
 \label{fig:steps-methodology}
\end{figure}

We propose a recommendation process as a two-step approach consisting of retrieval and ranking steps. In the retrieval step, recommendation candidates are determined by an existing RS. These candidates are ordered in the ranking step, and the top $N$ elements are returned to the user. Figure \ref{fig:steps-methodology} presents an overview of the recommendation process.

This investigation treats the RS algorithm as a closed system over which we can not exert any influence. Our solution emphasizes and contributes to the ranking stage. We determine an item's rank based on its impact on network metrics correlating with surprise. Section~\ref{sec:eval_procedure} presents a list of metrics investigated.

To evaluate such metrics, we construct KGs from datasets suited for the recommendation task (\textit{cf.} Section~\ref{sec:eval_datasets}).
Two types of KGs are constructed. The first type is a KG representing the catalog, \textit{i.e.}, containing the entire knowledge about the catalog. This includes the whole set of recommendable items and all the metadata describing them. The second type stands for the user-profile KGs. They are subgraphs of the catalog KG, representing items users have already interacted with. Such KGs are constructed considering TBox statements representing the domain of their datasets.  Therefore, including recommendable entities, as well as additional entities, along with heterogeneous relations.

\begin{figure}[!htp]
    \centering
    \includegraphics[width=0.7\linewidth]{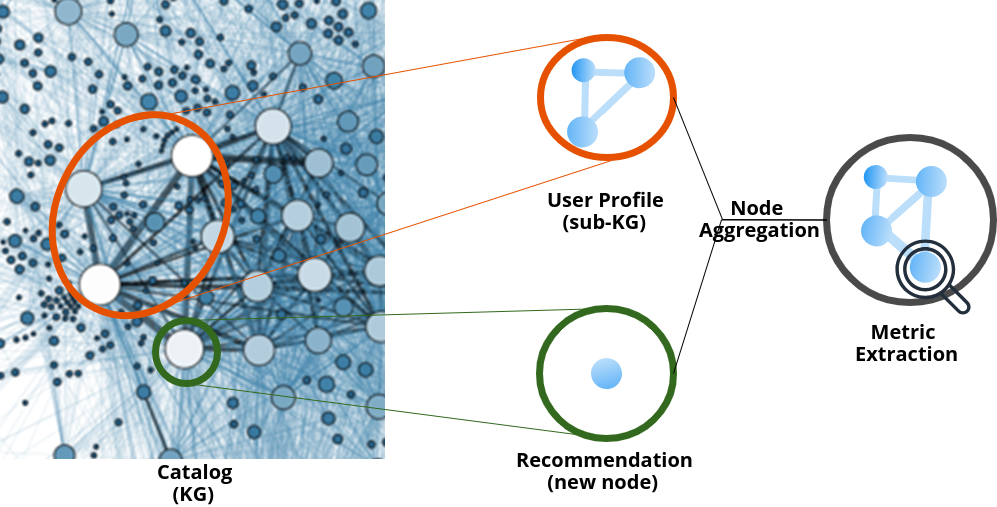}
    \caption{\textbf{KG-Informed Recommendation}. The sub-KG represents a user profile. A node representing a recommendation is obtained from the catalog KG and included in the sub-KG along with applicable edges. We then compute network metrics from the sub-KG.}
    \label{fig:method}
\end{figure}

The recommendable entities are evaluated by including them in the user-profile KGs. According to those existing in the catalog KG, a recommendable entity is included along with its relationships and further entities. From the updated user-profile KG, we compute complex network metrics (\textit{cf.} Figure~\ref{fig:method}). The process is conducted for all the recommendable items and all available metrics. At the final stage, our solution provides a re-ranked recommendation list sorted according to each metric.

Where network metrics do not result in scalar values, but in distributions, we calculate the Herfindahl-Hirshman-Index (HHI)~\cite{Hirshman1964} to obtain a single value representing the concentration of the network (\textit{cf.} ~\cite{Schoenfeld2021}). Let $s$ be the relative centrality score over all vertices, and $N$ the number of vertices, then the index and its normalized form are given by
\begin{align*}
    HHI & = \sum_{i=1}^{N}s^2 \\
    HHI^{*} & = \frac{HHI - 1/N}{1 - 1/N}
\end{align*}

Values of $HHI^*$ range in $[0,1]$, with $0.0$ corresponding to a balanced network with no monopolies and a value of $1.0$ indicating a strongly centralized network.

Formally, let $KG$ be a domain KG, consisting of concepts $C$ and relations $R$. A user profile $U = \{u_1, u_2, \dots, u_n\}$ is a subset $U \subset C$ of concepts a user has interacted with. Each user profile constitutes an induced subgraph $SG \subset KG$ containing the history items and further concepts and relations. 

For a recommender system \textit{RS}, let $I = \{i_1, i_2, \dots, i_n\}$ be the list of recommended concepts, ordered by a score assigned through \textit{RS}. Then, $N_{KG}[i_n]$ denotes the closed neighborhood of the vertex $i_n$ in $KG$ that corresponds to this recommendation. Let $SG'$ denote the induced subgraph produced by including $N_{KG}[i_n]$ in $SG$, \textit{i.e}., by adding a recommendation to the user subgraph.

Lastly, let $m$ denote any graph-metric, $m_{baseline}=m(SG)$ the value of this metric on the original user subgraph, and $m_{update} = m(SG')$ its value after incorporating the recommendation in the subgraph. Then, our method for re-ranking recommendations is as follows:

\begin{enumerate}
\def\labelenumi{\arabic{enumi}.}
\item
  Given a dataset, construct a KG $KG$;
\item
  Given a user profile $U$, determine the subgraph $SG$ of $KG$ that contains all items in the user's profile and all relations and intermediate entities;
\item
  Given a set of items in $U$, determine recommendations
  using any RS;
\item
  For each recommended item $i_n$, obtain the metric $m$ on the subgraph $SG'$ including $i_n$;
\item
  Re-rank all items according to their impact on the given computed metric;
\end{enumerate}

Algorithm~\ref{alg:method} provides a formal formulation of our solution.

\begin{algorithm}
\caption{KG-informed recommendation}
\label{alg:method}
\begin{algorithmic}[1]
\REQUIRE $KG$ \COMMENT{Catalog KG}
\REQUIRE $SG$ \COMMENT{User profile subgraph}
\REQUIRE $RS$ \COMMENT{Blackbox recommender system}
\REQUIRE $m$ \COMMENT{A graph-metric}
\STATE $v_{user\_nodes} \leftarrow SG.get\_all\_nodes()$
\STATE $I \leftarrow RS(v_{user\_nodes})$
\STATE $I' \leftarrow \emptyset$
\FORALL{$i \in I$}
    \STATE $SG' \leftarrow SG.copy()$ \COMMENT{Evaluate always on the same instance}
    \STATE $SG'.add\_node(i)$
    \STATE $edge\_list \leftarrow \emptyset$
    \FORALL{$neighbor \in KG.get\_neighbors(i)$}
        \IF{$neighbor \in v_{user\_nodes}$}
            \STATE $edges \leftarrow KG.get\_edges(i, neighbor)$ \COMMENT{May be more than one}
            \STATE $edge\_list.insert(edges)$
        \ENDIF
    \ENDFOR
    \STATE $SG'.add\_edges(edge\_list)$
    \STATE $metric\_value \leftarrow metric(i, SG')$
    \STATE $I'.insert(metric\_value, i)$
\ENDFOR
\STATE $kginformed\_recommendation\_list \leftarrow sort(I', metric\_value)$
\RETURN $kginformed\_recommendation\_list$
\end{algorithmic}
\end{algorithm}

%% file: sections/04_evaluation.tex
\section{Experimental Evaluation}
\label{sec:evaluation}

We evaluate our proposed approach on two distinct music- and movie-domain datasets. Proceeding according to our defined method (\textit{cf.} Section~\ref{sec:proposal}), we obtain re-ranked recommendation lists for a set of users. As we focus our attention on \enquote{surprisingness} of recommendations, rather than measuring precision/accuracy, we turn to \enquote{beyond accuracy}-metrics commonly used in the evaluation of surprise and serendipity in RS, such as novelty and diversity (\textit{cf.}~\cite{Castells2015,Ge2010}).

To back up the insights obtained in this sense, we further measure the agreement of the re-ranked recommendation lists with those generated through the SOTA RS, which we treat as ground truth for ``expectable'' recommendations.
Measuring normalized discounted cumulative gain (nDCG) on the two lists allows us to identify whether the re-ranked variant deviates from the expectable recommendations.

Our study addresses the following specific research questions:
\begin{description}

\item[RQ1]
  Which network-level metrics correlate with key surprise elements such as novelty, unexpectedness, and novelty in recommendations?

\item[RQ2]
  Can these metrics be used to introduce more surprise into state-of-the-art recommender systems?

\end{description}

\subsection{Datasets}
\label{sec:eval_datasets}

We report on data collection and curation for the two domains investigated.

\subsubsection{LastFM}
\label{sec:eval_datasets_lfm}

\paragraph{LFM-1b}

We base our analysis of recommendations for the music domain on the LFM-1b dataset~\cite{Schedl2016}, which we enrich with two further datasets: acoustic features for a selection of tracks (the \textit{CultMRS} dataset) curated by Zangerle \textit{et al.}~\cite{Zangerle2020}, and musical genres annotating a subset of tracks within LFM-1b, kindly provided by Schedl \textit{et al.}~\cite{Schedl2020}.
The acoustic features contained in the dataset were retrieved via the Spotify API\footnote{\url{https://developer.spotify.com/documentation/web-api/reference/get-several-audio-features}} and serve as content-based features describing the nature of a track. Examples for these features are a track's \textit{tempo}, or \textit{danceability}. After merging the three datasets, we are left with 379 million listening events (\textit{cf.} Table~\ref{tab:lfm1b-filtering}).

\input{tables/lfm-stats}

\paragraph{KG construction.}

From the merged LFM-1b dataset, we construct a KG consisting of artists, tracks, and genres. To model the relations among these entities, we use classes and properties provided by three different ontologies: FOAF\footnote{\url{http://xmlns.com/foaf/0.1/}}, Dublin Core\footnote{\url{http://purl.org/dc/elements/1.1/}} and Music Ontology\footnote{\url{http://purl.org/ontology/mo/}}~\cite{Raimond2007}; we define an auxiliary URI to identify entities from LastFM\footnote{http://last.fm/lfm-resource}. For instance, a description of the track \enquote{Never Gonna Give You Up} by Rick Astley in Turtle would be:

\begin{verbatim}
lfmr:disco a mo:Genre ;
    dc:title "disco" .
lfmr:15160 a mo:MusicArtist ;
    foaf:name "Rick Astley" .
lfmr:t_4471632 a mo:Track ;
    dc:title "Never Gonna Give You Up" ;
    mo:genre lfmr:disco ;
    foaf:maker lfmr:15160 .
\end{verbatim}

\paragraph{Recommendations.}

We sub-sample the listening events to 1000 users with at least 100 unique tracks in their profile. The mean number of tracks listened to is 1076 ($\pm 1194$), with a median of 656 tracks.
We use the Python library \textit{Surprise}~\cite{Hug2022}, which relies on explicit user-item-ratings to determine the base recommendations. As our data contains implicit ratings as the number of times a track was listened to by a user, we follow the approach outlined in Kowald~\textit{et al.}~\cite{Kowald2021} and scale these play-counts into the range $[1,1000]$ using min-max-normalization; a user's most-listened track will thus receive an explicit rating of 1000.

We evaluate six recommendation models provided by \textit{Surprise}: \textit{BaselineOnly}, which predicts a baseline rating estimate from global averages and user/item deviations (\textit{c.f.} Koren~\cite{Koren2010}); \textit{KNNBasic}, a user-based collaborative filtering approach using kNN; \textit{KNNBaseline}, \textit{KNNWithMeans}, and \textit{KNNWithZScore}, extensions of the base kNN model taking into account baselines, mean ratings, and z-score normalized ratings, respectively; and \textit{NMF}, a non-negative matrix factorization model.
We use the default parameters provided by the library and employ cosine similarity as the distance measure for the kNN-based approaches.

Using 5-fold cross-validation, we evaluate each algorithm's mean absolute error (MAE), and pick NMF as our final model; Table~\ref{tab:recos-mae} presents MAE for all models.

We train NMF on the full data and retrieve rating predictions on the \textit{anti-testset}, \textit{i.e.}, on all items present in the training data that the user has not rated. The recommendation lists obtained this way are truncated to the top 100 items, sorted by descending predicted rating.

\input{tables/tab-reco-eval}

\subsubsection{Netflix}
\label{sec:eval_datasets_nflx}

\paragraph{Netflix titles dataset}

Our evaluation includes the domain of movies and TV shows. For this matter, we considered the ``Netflix titles'' dataset, available on Kaggle\footnote{\url{https://www.kaggle.com/datasets/shivamb/netflix-shows}}. Such dataset provides a set of 8808 titles available on Netflix, considering movies and TV shows' titles, cast, directors, countries, release dates, ratings, and a brief description. All data is provided in a comma-separated value (CSV) file.

\paragraph{KG construction.}

The catalog KG was created considering TBox statements representing relationships and types as provided in the CSV file. Such TBox contains the type of each entry, which can be either a movie or a TV show. An actor acts on entries, and a director directs entries. Entries have an English title, a brief description, a country of origin, a rating, and a duration. All classes are of the \texttt{rdf:Class} type, and properties of the \texttt{rdf:Property} type.

The dataset provides no user data, therefore, we randomly generated 88 user-profiles, ranging from a minumum of 5 to 55 entries, representing watched movies and TV shows. From such, user-profile KGs were generated according to the same TBox as the catalog KG.

\paragraph{Recommendations.}

The recommendations were generated with the help of a state-of-the-art RS \cite{wang2019kgat}. We used the same parameters for the configuration of graph convolutaional layers, decay factors, and learning rates as the authors of the paper used for their evaluations.

The data set was cleaned up to obtain meaningful yet compact KGs. To do this, \texttt{rdf:label}-entities and nodes with a degree of $1$ were removed. In addition, the \texttt{rdf:Class} and \texttt{rdf:Property} nodes were removed to prevent the knowledge graph from becoming too centralized. Certain entries in the KGs were labeled as recommendable items, \textit{i.e.}, only movies and TV shows.

From user-profile KGs, the interactions on recommended items were registered and divided into training and test data using a 90/10 split, \textit{i.e.}, 90\% of the interactions of a usage profile were listed in the training set.

\subsection{Experimental Procedure}
\label{sec:eval_procedure}

For both datasets, we employed the following evaluation procedure:

\begin{enumerate}
    \item Obtain base recommendations through SOTA model.
    \item Re-rank base recommendations according to graph metric (as outlined in Section~\ref{sec:proposal}); ranking proceeds in ascending and descending order of item relevance.
    \item (LFM-1b only) For each metric and each sort order, measure Unexpectedness and Intra List Diversity using item features.
    \item For each metric and each sort order, compare the re-ranked with the base lists using nDCG@10 via TREC\_EVAL\footnote{\url{https://trec.nist.gov/trec_eval/} (As of Apr. 2024).}.
\end{enumerate}

The network metrics applied in this evaluation are the number of nodes, number of edges, density, PageRank, average degree, \{in,out\}-degree, betweenness, and closeness centrality.  
Such metrics adhere to standard metrics in the field of social network analysis \cite{Wasserman1994}.

To evaluate Unexpectedness and Intra List Diversity, we represent each track as an 8D vector of acoustic features.
The features we use are \textit{danceability}, \textit{energy}, \textit{speechiness}, \textit{acousticness}, \textit{instrumentalness}, \textit{liveness}, \textit{valence} and \textit{tempo}. In the original dataset, these features range in $[0,1]$, except for tempo, which we scale into this range using min-max normalization following prior research~\cite{Zangerle2020,Kowald2021}.

Intra List Diversity (ILD) measures the pairwise distance of all items in a recommendation list $I$ concerning a distance $d$ (c.f.~\cite{Castells2015}):

\begin{equation*}
    ILD(I) = \frac{1}{|I|\cdot(|I| - 1)}\sum_{i \in I}\sum_{j \in I}d(i, j)
\end{equation*}

We measure Unexpectedness on a user-profile level to determine how different a recommendation is from the user's previous history. Essentially, this is the mean distance of each new item to each item the user has interacted with. Thus, for a user-profile $H$, a recommendation list $I$ and a distance $d$, Unexpectedness can be expressed as (\textit{cf.} ~\cite{Castells2015}):

\begin{equation*}
    Unexpectedness(R) = \frac{1}{|I|\cdot|H|}\sum_{i \in I}\sum_{h \in H}d(i, h)
\end{equation*}

For both ILD and Unexpectedness, we employed cosine distance between feature vectors as $d$. In these measures and in nDCG, we limit the recommendation lists to the top $10$ items, in line with previous findings on users' searching behaviour~\cite{Jansen2000,Silverstein1999}.

To further assess the rank-based dynamics underlying this re-ordering, we measure nDCG@10 for each re-ranking.
The base recommendations serve as a ground truth of expectable recommendations for our purposes. Their ranking thus serves as the relevance judgement of items. High nDCG indicates that the same items are ranked highly in the base and re-ordered recommendations, whereas low nDCG indicates more perturbation in the second list. Our assumption here is that low nDCG indicates that highly expectable items are ranked lower after re-ordering.

\subsection{Experimental Results}
\label{sec:eval_results}

We present the results achieved from the experimental procedure for both datasets.

\subsubsection{LastFM.}

We first review the findings from measuring surprise on the re-ranked list of recommendations before evaluating list perturbation. Section~\ref{sec:discussion} discusses obtained results and how they can be further interpreted from a network perspective.

\begin{figure}
    \centering
    \begin{subfigure}[b]{0.49\textwidth}
        \centering
        \includegraphics[width=\textwidth]{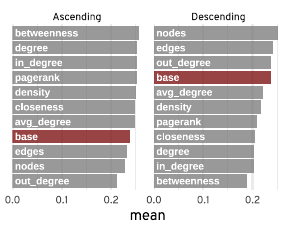}
        \caption{Unexpectedness}
        \label{fig:lastfm_unexp}
    \end{subfigure}
    \hfill
    \begin{subfigure}[b]{0.49\textwidth}
        \centering
        \includegraphics[width=\textwidth]{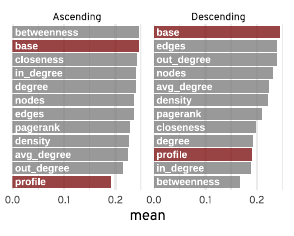}
        \caption{Diversity}
        \label{fig:lastfm_diversity}
    \end{subfigure}
    \caption{Measuring surprise on feature-level for recommendations re-ranked by metric. For Unexpectedness, the highlighted bars denote the comparison to the SOTA recommendations. For Diversity, the highlighted bars provide the intra-list values of SOTA and original user profiles (\textit{base} and \textit{profile}, resp.).}
    \label{fig:lastfm_surprise}
\end{figure}

We evaluate Unexpectedness and Intra List Diversity on all re-ranked recommendation lists and the two possible ranking orders.
We include the measurements obtained on the original SOTA recommendations as a baseline; in addition, the users' mean profile diversity serves as a reference point for diversity.
Figure~\ref{fig:lastfm_surprise} plots the mean measures against all metrics, split by ranking order. For Unexpectedness, sorting ascendingly by betweenness results in the largest deviations from the user's history, as shown in Figure~\ref{fig:lastfm_unexp}.

\enquote{Betweenness} in this case corresponds to the Herfindahl-Hirshman-Index (HHI) of the distribution. Sorting in ascending order thus places low index values at the top of the list, indicating a fairer distribution and therefore an overall less centralized network. The opposite holds true for descending order, where favouring high betweenness-indexes and thus more centralized networks results more expectable recommendations. We observe that increasing the number of nodes and edges in the users' subgraphs has the highest effect on Unexpectedness.

Turning to Diversity, we first observed that users' listening behavior seems largely uniform, as indicated by the \textit{profile} bars in Figure~\ref{fig:lastfm_diversity}.
As the measure of ILD on the user profile is expressed as the mean pairwise distance between items in the list, a low distance on average indicates the presence of tracks with similar acoustic features.

We found that preferring a low betweenness index and thus decentralized networks results in diverse tracks being recommended for ascending sort order, whereas for the descending case, increasing the number of nodes, edges, and out-degree produces the most diverse lists out of our approaches but does not outperform the baseline recommendations. An interesting observation is that the base recommendations already contain very diverse items. This is in line with previous findings of the underlying algorithm, NMF, being able to recommend items from the long-tail of user-item-interactions~\cite{Kowald2020}.

\begin{figure}
    \centering
    \includegraphics[width=\textwidth]{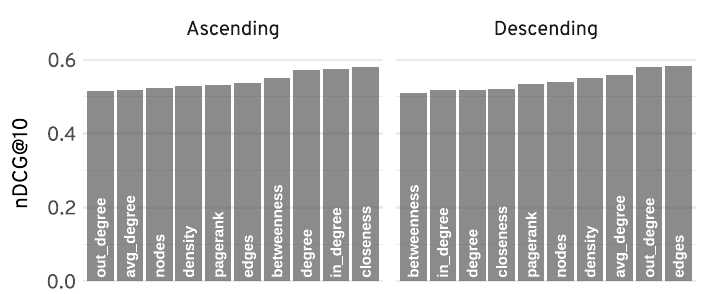}
    \caption{\textbf{nDCG@10} for LastFM. Panels show nDCG scores obtained on recommendation lists ranked by metric in ascending and descending order.}
    \label{fig:lastfm_ndcg}
\end{figure}

To assess the extent to which re-ranked lists correspond with the original, expectable ranking, Figure~\ref{fig:lastfm_ndcg} plots nDCG@10 for all metrics. We found that out-degree and betweenness, particularly, result in a high perturbation of ranks. We highlight that items considered relevant by an expectable RS are not ranked highly after optimizing for one of the network metrics.

\subsubsection{Netflix.}

\begin{figure}
    \centering
    \includegraphics[width=\textwidth]{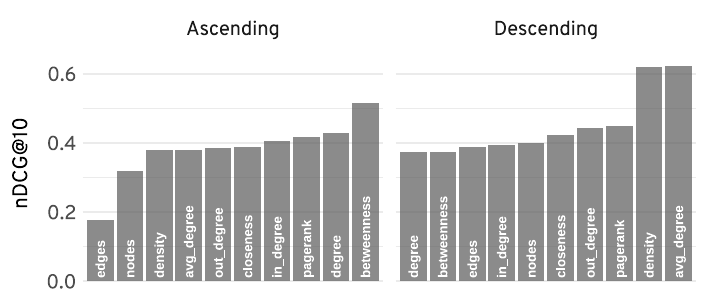}
    \caption{\textbf{nDCG@10} for Netflix. Panels show nDCG scores obtained on recommendation lists ranked by metric in ascending and descending order.}
    \label{fig:netflix_ndcg}
\end{figure}

Unlike LastFM, Netflix titles dataset provides no additional features (such as acoustic features) to enable performing observations based on Unexpectedness and Diversity. The analysis on this dataset is entirely based in the nDCG score, considering the initial 10 elements of the recommendation list. Figure~\ref{fig:netflix_ndcg} summarizes such results by metrics, both in ascending and descending orders of relevance.

Similarly to LastFM, we observe that betweenness centrality is the metric that most introduces surprise into the list of recommendations obtained from the RS. To further illustrate this finding, Table~\ref{tab:results} presents the first three recommendations offered by the RS, compared with those re-ranked through betweenness centrality in particular user-profile KGs. Table~\ref{tab:results} illustrates a sample for each dataset.

\input{tables/eval_list_comparison}

%% file: tables/lfm-stats.tex
\begin{table}
\centering
\caption{Statistics of LFM-1b after merging with other datasets.}\label{tab:lfm1b-filtering}
\begin{tabular}{lr}
\toprule
\# listening events &  379.754.730\\
\# users & 120.053\\
\# artists & 26.129 \\
\# tracks & 282.011\\
\# genres & 2.137\\
\bottomrule
\end{tabular}
\end{table}

%% file: tables/tab-reco-eval.tex
\begin{table}
\centering
\caption{Evaluation of prediction algorithms.}\label{tab:recos-mae}
\begin{tabular}{lr}
\toprule
Model & MAE \\
\midrule
\textbf{NMF} & \textbf{54.82} \\
BaselineOnly & 62.38 \\
KNNWithZScore & 65.74 \\
KNNBaseline & 67.01 \\
KNNWithMeans & 67.47 \\     
KNNBasic & 71.1 \\
\bottomrule
\end{tabular}  
\end{table}

%% file: tables/eval_list_comparison.tex
\begin{table}[!htp]
\caption{Comparing the top three recommended items obtained from state-of-the-art recommender, against those re-ranked using betweenness centrality applied on a user-profile KG.}\label{tab:results}
\resizebox{\textwidth}{!}{%
\begin{tabular}{@{}llll@{}}
\toprule
Dataset                  & SOTA recommender                      & Re-ranked (betweenness)             &  \\
\midrule
\multirow{3}{*}{Netflix} & Bakugan: Armored Alliance             & Creeped Out                         &  \\
                         & The C Word                            & Black Mirror                        &  \\
                         & Weird Wonders of the World            & Arthur Christmas                    &  \\
\midrule
\multirow{3}{*}{LFM-1b}  & Iron Maiden, The Talisman             & Shakira, Spotlight                  &  \\
                         & Iron Maiden, When the Wild Wind Blows & Here We Go Magic, Make Up Your Mind &  \\
                         & Shakira, Spotlight                    & Here We Go Magic, Alone But Moving  & \\
\bottomrule                    
\end{tabular}%
} 
\end{table}

%% file: sections/05_discussion.tex
\section{Discussion}
\label{sec:discussion}

We observe that recommendations sorted by betweenness in ascending order of the associated HHI exhibit high Unexpectedness and Diversity. Ranking in this way favors nodes that result in a lower HHI, thus, revealing a more decentralized user subgraph. In such a KG, many paths among concepts exist, and there is low monopolization. The opposite holds for a highly centralized KG, in which a small number of concepts appear along many paths and carry high importance.

\begin{figure}
    \centering
    \begin{subfigure}{.32\textwidth}
        \centering
        \includegraphics[width=\linewidth]{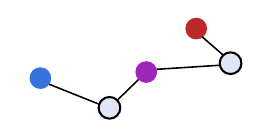}  
        \caption{Profile subgraph...}
        \label{fig:network_betweenness:profile}
    \end{subfigure}
    \begin{subfigure}{.32\textwidth}
        \centering
        \includegraphics[width=\linewidth]{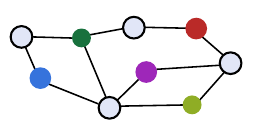}  
        \caption{...with diverse...}
        \label{fig:network_betweenness:democratic}
    \end{subfigure}
    \begin{subfigure}{.32\textwidth}
        \centering
        \includegraphics[width=\linewidth]{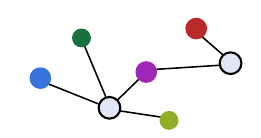}  
        \caption{...and similar items}
        \label{fig:network_betweenness:highbetweenness}
    \end{subfigure}

    \caption{Illustration of the effect of recommendations on betweenness centrality in profile subgraphs. Diverse items being recommended open alternative paths in the resulting profile subgraph lowering the betweenness of all nodes (\ref{fig:network_betweenness:democratic}), whereas similar items tend to increase the betweenness of a few nodes (\ref{fig:network_betweenness:highbetweenness})}
    \label{fig:network_betweenness}
\end{figure}

Figure~\ref{fig:network_betweenness} illustrates this effect, presenting an example user subgraph and two extensions arising from incorporating more diverse (\textit{cf.} Figure~\ref{fig:network_betweenness:democratic}) or more similar (\textit{cf.} Figure~\ref{fig:network_betweenness:highbetweenness}) recommendations. Interactions and recommendations are shown as solid colored circles; related concepts are light colors with an outline. Diverse items will likely be loosely connected to existing concepts the user is familiar with and bring along further related nodes, thus expanding the user's exposure. Contrast this with the second example, where similar items are introduced that only exhibit relations to concepts familiar to the user. These examples illustrate the effect on the number of edges, nodes, and degree-related measures. In the diverse case, adding two recommendations results in four nodes and seven edges added to the graph versus two nodes and two edges for the case of similar items.

Considering the results from evaluating nDCG, we observed that ranking by betweenness, node-/edge counts, or degree-based metrics yields lists with low-rank correlation compared to expectable recommendations. 

Our study demonstrated that network-level metrics correlate with key surprise elements such as diversity and unexpectedness (\textbf{RQ1}). We found betweenness resulting in the most diverse and unexpected recommendations that rank expectable items lower than a state-of-the-art baseline. We showed that adding a KG-informed re-ranking model on top of an existing recommender system can thus introduce a level of surprise into user-item-recommendations (\textbf{RQ2}).

Results highlighted that calculating betweenness may not be computationally feasible in constrained environments, especially on large profile subgraphs. Besides truncating user profiles to the most recent interactions as a solution in this case, our findings suggest that node-/edge counts or degree-based features are viable alternatives to betweenness.

We identify the Netflix dataset's lack of rich content-based features, prohibiting a similar investigation of surprise-related measures as performed for the enriched LFM-1b dataset. Furthermore, a user study should evaluate the degree of surprise, as listening and viewing behaviors are governed by highly subjective user dynamics. We plan to address it in future studies.

The catalog KGs employed in our study only contain intra-domain concepts (artists, music genres, directors, actors, etc.). However, KGs are well suited for linking cross-domain concepts, \textit{e.g.}, tracks that appear in a movie's score, or actors who are musicians. Not only does this result in a richer representation of domains, it also enables cross-domain recommendations. We defer an analysis of surprising recommendations in such settings to future work.

%% file: sections/06_conclusion.tex
\section{Conclusion}
\label{sec:conclusion}

We still encounter open research challenges in how systems may deal with and benefit from surprise recommendations. This investigation designed a solution incorporating network-level metrics to introduce personalized yet unexpected recommendations to users. We evaluated the LastFM music and Netflix movies datasets to determine the extent of intra-list diversity, unexpectedness, and comparison to nDCG, respectively, affect the degree of surprise in recommendations. We found that network-level metrics indeed influence the degree of surprise in recommendations. Our results demonstrated that betweenness centrality showed a stronger influence when reranking recommendations for a surprise. Future work involves additional analysis of surprising recommendations and how textual features from items can be combined with our designed approach.